# Physical Questions Posed by DNA Condensation

Bae -Yeun Ha* and Andrea J. Liu
Department of Chemistry and Biochemistry
University of California, Los Angeles, CA 90095

*Current address: Department of Physics, Simon Fraser University, Burnaby, B. C. V51 1S6, Canada

## ABSTRACT

When multivalent salts are added to dilute DNA solutions, the DNA chains condense into bundles of a well-defined size. We explore two physical problems motivated by this phenomenon: the origin of the attractive interactions between the highly negatively-charged DNA chains that cause them to concentrate into bundles, and the mechanism that prevents the bundles from growing indefinitely large.

## I. INTRODUCTION

DNA in aqueous solution carries a very high negative charge, with two electronic charges per base pair (or equivalently, per 3.4 Å) along the length of the double helix. In its unpackaged form, DNA is well-described as a wormlike chain with a persistence length of approximately 500 Å [1]. However, when DNA is packaged in viruses and cells, it is highly concentrated into configurations where helices are approximately parallel to each other and separated by roughly 5 Å of water. For example, the DNA of the T7 bacteriophage is approximately $10^{-4}$ times smaller in the phage head than in its unpackaged form [2]. To concentrate DNA to this level by brute force would require a pressure of over 100 atm [3].

It has been suggested that many bacteriophages [4-7] use multivalent cations to package their DNA. When multivalent cationic species (polyamines) known to exist in host bacteria are added to DNA in dilute solution, the chains spontaneously form highly-concentrated toroids, of the same shape and size as the *in vivo* packaged DNA. This phenomenon is known as DNA condensation [8, 9]. In addition to polyamines typically found in cells, a wide variety of cations can condense DNA *in vitro*, generally requiring only that the valency of the cation is three or higher [8, 9]. Moreover, it turns out that DNA is not the only polyelectrolyte able to condense in this way. Experiments on other stiff, highly-charged polyelectrolytes such as F-actin [10], tobacco mosaic virus and fd virus [11], also condense into dense bundles. The fact that the attractions are observed for a wide range of stiff polyelectrolytes and a variety of multivalent counterions indicates that specific interactions are not responsible, and that the mechanism may be susceptible to a general electrostatic analysis.

Another striking experimental feature is that the attractions do not appear to lead to macroscopic phase separation. In this sense, the counterion-mediated attraction between the chains appears to have a different character from ordinary attractions that lead simply to phase separation at sufficiently high concentrations. Instead, the chains tend to form dense bundles of a fairly well-defined thickness [8, 11]. The precise morphology of the bundles appears to depend sensitively on the persistence length of the polyelectrolyte, the chain length and the concentration. In the case of dilute DNA, the bundles tend to be toroidal or rod-shaped. Other stiff polyelectrolytes tend to form rodlike bundles or networks of bundles. In each case, however, there is a well-defined cross-sectional thickness for the bundles. We will concentrate on the question of why there is a characteristic cross-sectional bundle diameter, rather than on the specific morphology of the bundles.

## II. ORIGIN OF COUNTERION-MEDIATED ATTRACTION BETWEEN LIKE-CHARGED RODS

The question of how counterions mediate attractions between like-charged chains has garnered much attention in the physics community because mean-field Poisson-Boltzmann theory predicts that like-charged rods should repel, regardless of the valency of the counterion [12]. This implies that fluctuation and correlation effects that are neglected in Poisson-Boltzmann theory must be responsible for the attraction.

### A. COUNTERION-MEDIATED ATTRACTION BETWEEN LIKE-CHARGED PLATES

The mechanism for attraction has also been addressed for a different geometry, namely two like-charged parallel plates in solution. There, integral equation methods [13, 14] show that fluctuations within the hypernetted chain approximation (but neglected within mean-field theory) could lead to attractions. Density functional arguments also predict that like-charged plates can attract [15]. Complementary to these two liquid-state approaches is a low-temperature picture, where the counterions crystallize in the plane parallel to the plates [16]. It is obvious that the plates should attract in this ionic crystal phase. The precise form of the attraction depends on fluctuations of the ions around their lattice positions (phonon modes) [17], but clearly the attraction should be strongest at zero temperature. Finally, field theoretical approaches based on one-loop corrections to either Debye-Hückel [18] or Poisson-Boltzmann [19] theory also show that two plates can attract. These theories are appropriate to the high-temperature noncrystalline phase, where thermal fluctuations in the charge density along the plane become correlated and give rise to an attraction similar to the van der Waals attraction. This attraction vanishes at zero temperature.

### B. COUNTERION-MEDIATED ATTRACTION BETWEEN LIKE-CHARGED RODS

There is an important distinction between the plate geometry (the interaction between two like-charged planes) and the rod geometry (the interaction between two like-charged rods), in that the origin of the repulsion is completely different in the two cases. In the plane geometry, all the counterions are condensed. As a result, there is no net effective charge on the plates. On the other hand, the confinement of counterions between the plates leads to an entropic repulsion; this double-layer repulsion is generally much larger than the electrostatic contribution.

For the rod geometry, on the other hand, only a fraction of the backbone charges are neutralized by condensed counterions. This means that each rod still carries a net charge, so two rods will repel each other electrostatically. On the other hand, the entropic repulsion due to confinement of counterions is much weaker.

The fact that the origin of the repulsion is different for plates and for rods means that the overall phase behavior can be different, because the interaction depends sensitively on the competition between the correlation-attraction and the repulsion. However, the mechanisms of attraction that have been proposed for rods are the same as for plates. Thus, the ionic crystal picture for plates has been applied to rods [20-23], as has the thermal fluctuation picture (which was developed for rods more than 30 years ago by Oosawa [24-26]). In the case of rods, there are several versions of the ionic crystal model, which differ in microscopic details [21-23, 27]. In the thermal fluctuation picture, fluctuations in the condensed-counterion density along the rods lead to nonuniformities in the charge distribution, which can become correlated from one rod to another [26, 28], leading to an attraction similar to the van der Waals interaction. We have introduced a third approach, called the charge-fluctuation approach [29, 30], which is an extension of the thermal fluctuation approach to ions of nonzero size, and which captures aspects of both the thermal-fluctuation picture and the ionic-crystal pictures.

## C. LOW-TEMPERATURE IONIC CRYSTAL PICTURE VS. HIGH-TEMPERATURE THERMAL FLUCTUATION PICTURE

The discrepancy between the “ionic crystal approach” and the “thermal-fluctuation” approach described above has led to considerable controversy in the rod case [31, 32]. We emphasize that there is no way to reconcile the two approaches completely, because the thermal-fluctuation approach is a high-temperature approximation, while the ionic-crystal approach is a low-temperature approximation, and the two are separated by a phase transition, namely crystallization. The same controversy has arisen in other contexts, such as one-component and two-component plasmas [33-35]. The thermal fluctuation approach neglects correlations that are important at lower temperatures because it neglects a length scale that is important at low temperatures or high densities, namely the ionic size. This size prevents the attractive interaction from diverging and is clearly important to ionic crystallization because if all the ions were point charges, the system would collapse onto a point at zero temperature. Most field-theoretical calculations have neglected this size by assuming point ions. However, we have shown that it is straightforward to include this length scale by assuming that charge is correlated over the size of the ion [30]. We call this approach the charge-fluctuation approach to distinguish it from the thermal-fluctuation approach, which assumes point ions. When the nonzero ionic size is incorporated in this way, we find oscillatory charge correlations that grow in range as the temperature is lowered, and eventually diverge at the spinodal for the ionic crystal [29]. Note that it has previously been shown for electrolyte solutions that full Debye-Hückel theory, which includes the nonzero ionic size (as opposed to the Debye-Hückel limiting law, which assumes point ions), also produces oscillatory charge correlations indicative of an incipient ionic crystal [36]. Thus, the one-loop field-theoretical approach leads to reasonable qualitative behavior at low temperatures, even though it is unlikely to be quantitatively accurate there. More recently, it has been shown for the case of two plates [37] that at high temperatures, long wavelength fluctuations in the charge density along the plates dominate the interaction, while at low temperatures, short wavelength fluctuations dominate. The long wavelength fluctuations lead to an attraction that diminishes as temperature is lowered, while the short wavelength fluctuations lead to an attraction that increases as temperature is lowered. It is known from simulations that the attraction strengthens as the temperature decreases [20]; this implies that the short wavelength fluctuations are more important. The short wavelength fluctuations in the charge-fluctuation model represent liquidlike correlations that are the analogue of the ionic crystal correlations below the freezing transition. Thus, we argue that at higher temperatures where the charge correlations are liquidlike, a charge fluctuation picture might more accurately describe the origin of the attraction, but at lower temperatures where the charge correlations are solidlike, the ionic crystal is a better description. Certainly, our approach is a poor one in the ionic crystal phase, just as the ionic-crystal approach is a poor approximation above the melting transition. Simulations show that the interactions between chains can be strongly attractive even when there is significant counterion diffusion along the rods, and only liquidlike ordering of the condensed counterions [20, 38]. This is the regime in which the charge-fluctuation picture is appropriate.

## D. CHARGE-FLUCTUATION MODEL

Our calculations have been discussed extensively elsewhere [30], but it is worthwhile to describe the assumptions underlying the model and approach. We start with rods with a uniform negative charge. The actual charge distribution for DNA is helical; this nonuniform distribution can also lead to attractions at very short distances and low temperatures [39], but we have not adopted this more realistic description. The charge on the rods is balanced by counterions, and we also allow for salt. In reality, the counterions are distributed with some spatial density profile around the rods, which can be approximated by the solution to the nonlinear Poisson-Boltzmann equation. We adopt the two-state approximation to describe this density distribution; that is, we divide the counterions into two classes, condensed and free [24, 40]. A condensed counterion is approximated to lie on the nearest monomer, and to add a charge $Z$ to the net charge of that monomer, while a free counterion contributes to Debye screening of the electrostatic interactions in the solution. Because condensed counterions can move along the rods or exchange with free counterions, the effective charge of a monomer can fluctuate. If we assume that a large number of condensed counterions can be assigned to a given monomer, then we can apply the central limit theorem to the charge distribution, and can treat the charge per monomer as a Gaussian variable. Thus, we characterize the charge distribution by two quantities: the net charge per monomer, $q = -f_0 + Zf_c$, and the variance in the charge of a monomer on the rod, $\Delta = Z^2 f_c$. Here, $-f_0$ is the bare charge per monomer in units of the electronic charge, and $f_c$ is the fraction of condensed counterions per monomer.

Given the model with random charges on each monomer, we start with the Hamiltonian given by Coulomb interactions between all pairs of charges in the system. We assign charges on the rods a nonzero size, given by the counterion diameter. Charges on free ions in solution, however, are treated as point ions. (The point-ion approximation is a good one as long as the charge density is low. Under the conditions of interest for DNA condensation, the charge density along the rods is high due to counterion condensation, but the charge density in solution is low.) We integrate over free ions at the Gaussian level; that is, we assume that fluctuations in the density of mobile ions are small. In fact, it is possible to integrate over free ions exactly if they are point charges. We make this approximation in order to treat the mobile charges and the fixed charges on the rods in a consistent way. Tracing over the free ion positions at Gaussian order is equivalent to Debye-Hückel theory [26], while the exact point-ion integration is equivalent to Poisson-Boltzmann theory [41].

Finally, we integrate over the random charges on the rods at Gaussian order. This leads to a one-dimensional Debye-Hückel theory for the monomeric charges. However, it differs from a standard one-dimensional theory in two important respects. First, the charges on the rods have a nonzero size. In our Gaussian approach, the size enters via a form factor for the charged particle. Second, the charges interact with each other, and with the charges on other rods, via three-dimensional screened interactions. In other words, our approximation is equivalent to treating each rod as a one-dimensional Debye-Hückel system that interacts with itself and with all the other rods via a three-dimensional Debye-Hückel system of mobile ions.

How good are our approximations? We have made three main approximations. The first is to include counterion condensation within a two-state model. The second is to introduce the nonzero size within a form factor for the charged particle. The third approximation is to treat the fluctuations along the rods within a quasi-one-dimensional Debye-Hückel theory. Recent calculations by Kardar and Golestanian[19] avoid the two-state assumption by expanding around the Poisson-Boltzmann result. Their calculation provides some justification for the two-state model: when the counterion distribution is approximated with a step function, our expressions $q = -f_0 + Zf_c$ and $\Delta = Z^2 f_c$ are recovered [42]. Expanding around the Poisson-Boltzmann solution with its spatial distribution of counterions (without assuming a step function) is a definite improvement over our theory, but this approach is quite complicated to apply to our bundle system, especially if the ions are assigned a nonzero size.

The second approximation involving the ionic size can be improved upon even within the Gaussian approximation by including the structure of the ion in a self-consistent way [43]. It would be worthwhile to extend our calculations to include this approach.

The third approximation that we have adopted (describing the rods with associated condensed counterions as one-dimensional Debye-Hückel systems coupled to each other through a three-dimensional Debye-Hückel ionic solution) relies on the first term in a perturbation (loop) expansion that is best at higher temperatures and probably diverges in the regime of interest. However, it is important to recognize that good approximations can be useful beyond their range of validity. The important question is: what physics is left out by our description? In the case of *simple* Debye-Hückel theory, two important qualitative effects are left out: ionic associations (counterion condensation) and the possibility of oscillatory charge correlations. We have gone beyond the simple theory by including counterion condensation within a two-state model. We have also extended the simple theory to include oscillatory charge correlations by allowing for a nonzero ionic size (this is described in greater detail in Section II.D). Thus, although our approximations will not lead to *quantitatively* accurate behavior at low temperatures, they predict the correct *qualitative* behavior.

Our approach is similar in spirit (although different in form) to one adopted by Fisher and coworkers to describe criticality in electrolyte solutions. They also adopt a “two-state” model to include Bjerrum pairs in chemical equilibrium with free ions, and they note that it is essential to include the nonzero ionic size in order to capture criticality [35].

Finally, we note that the model itself is not accurate in the ionic crystal phase. The condensed counterions sit *on* the rods in our model, whereas they really should sit in between the rods. In other words, the structure of the ionic crystal is not captured correctly by the model. Recent calculations argue that multipole moments perpendicular to the rod axis that arise when counterions lie *between* rods are important to the low-temperature behavior [23]. This is probably a source of greater quantitative error at low temperatures than is the Gaussian approximation.

# III. ORIGIN OF CHARACTERISTIC CROSS-SECTIONAL DIAMETER OF BUNDLES

Regardless of whether the ionic crystal or charge-fluctuation picture is adopted, the same picture emerges for the interaction between two parallel rods. In both pictures, the range of the attraction is set by the characteristic length scale of charge fluctuations along the length of the rods. Most of the backbone charges are neutralized by condensed counterions in the regime of interest to DNA condensation, so the characteristic length scale is roughly the distance between condensed counterions along the backbone, namely half a base pair. This length scale is extremely short, so the range of the attraction is also short. There is also a repulsion due to the net-charge on each rod, whose range is set by the screening length. Under conditions where DNA condenses, the repulsion is longer in range than the attraction. Many other systems have short-ranged attractions and longer-ranged repulsions [44]. In all those cases, the competition between the two interactions leads to a finite domain size. One would therefore expect the counterion-mediated attraction to lead to stable finite-sized bundles, where the size of the bundle is limited by the range of the repulsion, namely the screening length. In fact, this argument is completely wrong. This can be seen from the experimental observation that the bundle size is much larger than the screening length. Below, we argue that the expectation of a stable finite bundle size is fundamentally flawed because it relies on the incorrect assumption that the counterion-mediated interactions are pairwise additive. Instead, the bundle should be viewed as a domain of a new concentrated phase, and the surface of the bundle is the interface between this concentrated phase and a dilute phase [45].

## A. NON-PAIRWISE-ADDITIVITY OF COUNTERION-MEDIATED INTERACTION

Neither the short-ranged attraction nor the long-ranged repulsion are pairwise-additive in the regime of interest [46]. First consider the attractive interactions. These arise from correlations between the charge density distributions on different rods. Suppose there are two rods with correlated charge distributions. If a third rod is introduced, its charge distribution will affect those on the other two rods, leading to a three-body contribution to the attraction. Thus, the interaction among the three rods must be computed explicitly, without assuming pairwise additivity.

Now consider the repulsive interactions. These originate from the net charge on each rod. That is, not all the backbone charges are neutralized by condensed counterions because the temperature is not zero, so each rod still carries a net negative charge. The net charge of the bundle (i.e. the sum of the net charges of the rods in the bundle) is therefore nonzero for all bundle sizes. However, the total charge enclosed in the bundle is not the same as the net charge of the bundle. The remaining counterions (the free counterions) partition themselves between the volume occupied by the bundle and the rest of the solution. As the bundle grows and occupies a larger fraction of the available volume, a larger fraction of the free counterions will be found in the bundle. In the limit of a very large bundle, the total charge enclosed in the volume of a bundle should be proportional to the surface area of the bundle.

In reality, we expect an even simpler picture. The dielectric constant of the chains is much lower than that of water. Inside a bundle, the effective dielectric constant is therefore quite low, because the chain concentration is extremely high. As a result, there should be nearly complete counterion condensation inside the bundle, so again the total charge of the bundle should be proportional to the surface area of the bundle.

We have calculated the energy of an $N$-rod bundle explicitly, using the one-loop approximation (the same approximation that we used for two rods) [30, 46]. Because the repulsion weakens as the bundle size grows, we find that the repulsion does not limit the bundle size. Instead, the free energy decreases linearly with $N$, the number of rods in the bundle. Thus, the counterion-mediated attraction is no different from more pedestrian attractions such as the dispersion attraction. This calculation does not include the translational entropy of the rods. Once the entropy is included, the equilibrium phase behavior of the solution will be just like any other solution of particles interacting via attractive interactions. At low concentrations the chains will remain in solution and at higher concentrations they will precipitate out to form a chain-rich phase in equilibrium with a chain-poor phase.

## B. KINETICS OF BUNDLE FORMATION

The analysis above appears to contradict the experimental finding of a well-defined finite bundle size. We argue that the experimental observation is not an equilibrium effect, and that the kinetics of bundle formation lead to a well-defined size. In fact, we believe that the competition of the short-ranged attraction and longer-ranged repulsion leads to unique kinetics of bundle growth [47].

In order to understand the kinetics of bundle growth, we must first consider the interaction between two rods as a function of the angle between them. The interactions between anisotropic particles such as rods depend sensitively on relative orientation. For most types of interactions, such as excluded volume or dispersion interactions, the interactions are anisotropic only in magnitude. The polyvalent-counterion-mediated interaction between like-charged rods is unusual because not only its magnitude but also its *sign* depends sensitively on the angle between the rods. This can be seen from the following argument. At large separations, two chains repel each other at all angles because the repulsion is longer in range than the attraction. At short separations, however, the two chains can attract each other if they are sufficiently parallel that their charge distributions can become correlated. If the chains are tilted away from parallel, less correlation arises, so the attraction decreases and the free energy increases. On the other hand, if the chains are perpendicular to each other, there is no correlation and the interaction is repulsive. As they are tilted away from a perpendicular configuration, the free energy increases because the repulsion increases; accordingly, there is a barrier in the free energy as a function of angle. This barrier has been calculated by us for two rods within the charge-fluctuation model [47], as shown in Fig. 1.

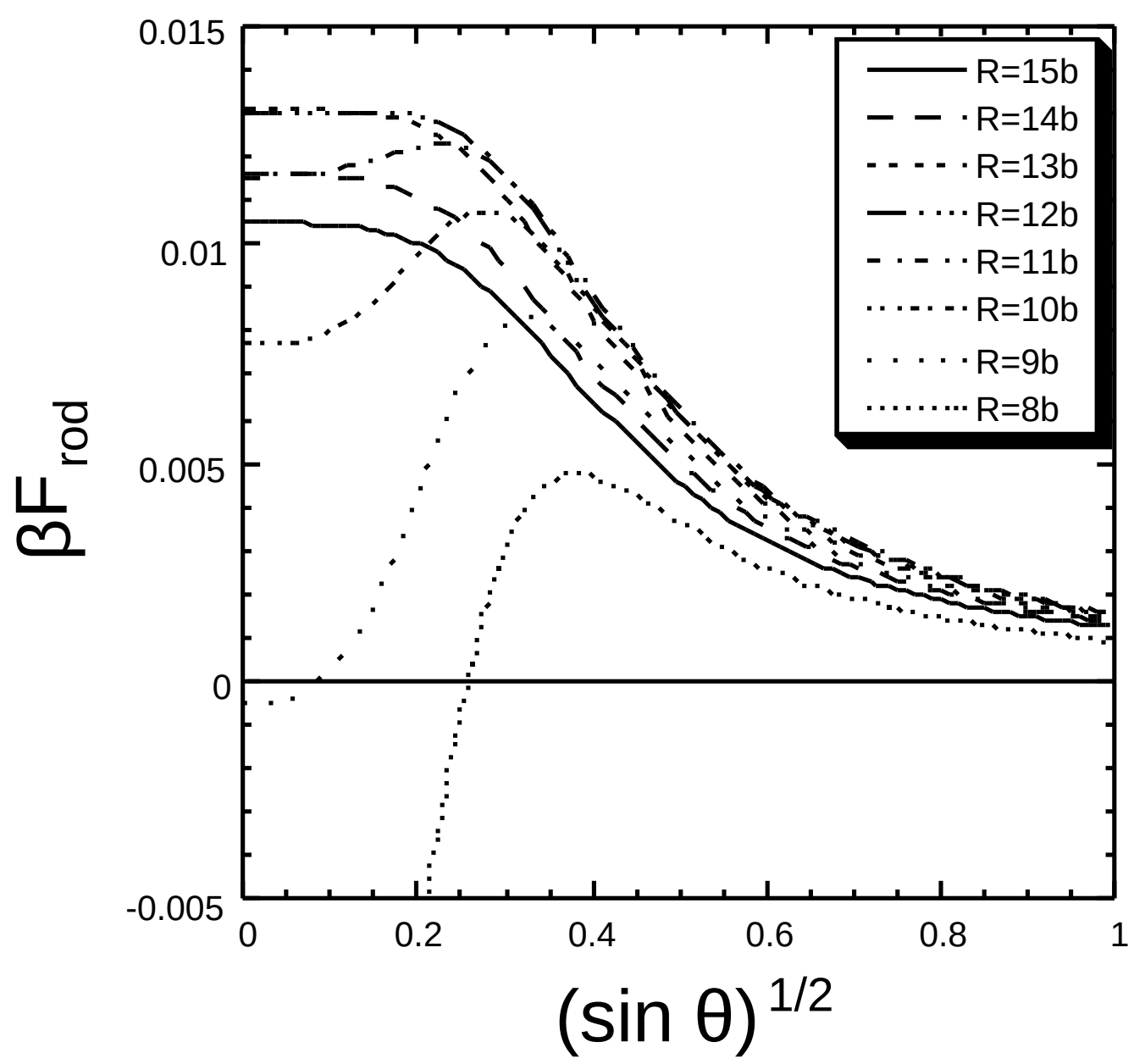

1. Plot of free energy of polyvalent-counterion-mediated interaction between two rods as a function of the angle between the rods for different rod separations, R (in units of the monomer size b).. The two rods are parallel at $\theta$=0. We then tilt one rod around the axis of separation between the two rods to an angle $\theta$. For this calculation, T=300K, $\varepsilon = 80$ and Z=2. The rod length is 500 Å (the persistence length of DNA). From Ref. [47].

Note from Fig. 1 that the barrier height is much smaller than the thermal energy for two rods. This is an artifact of our model, where we have modeled the rods as line charges with zero diameter. For more realistic charge distributions, e.g., spread out on the surface of a finite-diameter rod, we calculate the height to be tens of times larger [48]. The reason why the barrier height increases with rod diameter can be seen from a simple geometrical argument. Consider the circular cross-sections of two rods separated by some distance. Since the attraction is short-ranged, only the charge distributions along the facing arcs of the two circles can attract. On the other hand, the repulsion is longer ranged so charges at all points along the circular cross-sections can repel. As a result, the repulsion increases relative to the attraction with increasing rod diameter, leading to a higher repulsive barrier. For rods of the thickness of DNA, we find that the barrier is therefore significant, even for just two rods. Of course, since the attraction is short-ranged, it is very sensitive to microscopic details and our estimates should not be taken too seriously. However, the predicted *trends* are reliable and have interesting consequences for the kinetics of bundle growth.

To see how the barrier shown in Fig. 1 affects bundle growth, consider the interaction free energy between a chain and a bundle of already-condensed chains. This free energy contains a barrier as a function of the angle between the chain and the bundle. For sufficiently small bundles, a chain can overcome this barrier and join the bundle; however, the barrier height increases linearly with the bundle diameter because a skewed chain is repelled by the rods in the facing surface of the bundle. For a sufficiently large bundle, the barrier can exceed the thermal energy, but if a chain is very close and nearly aligned with a thick bundle it can rotate into a parallel orientation and join the bundle because the interaction is attractive. On the other hand, if a chain encounters the bundle at too large an angle, it will rotate into a perpendicular orientation to lower the free energy and thereby move away, because the interaction is repulsive. As the bundle grows larger, it is less and less likely that a chain will be able to approach the bundle at a small enough angle for it to fall into the attractive minimum.

Now consider what happens when a bundle of N chains encounters a bundle of M chains. Again, the charges on the facing surfaces of the two bundles will repel each other, leading to a barrier that increases with the product of their diameters. Thus, bundle-bundle aggregation is also discouraged as the bundles increase in size.

This kinetic mechanism is, to our knowledge, unique. In equilibrium, the system would phase separate into a dilute phase of randomly oriented chains and a concentrated phase of parallel, hexagonally ordered chains [49]. In most systems that phase separate, there is a nucleation barrier but no barrier to subsequent growth. In our case, the nucleation barrier is negligible [45]. (Note, however, that in the specific case of toroid formation there is a different nucleation barrier to forming a loop of the toroid [50].) Regardless of the nucleation process, there is a barrier to domain growth that increases in height as the domains grow. This inevitably prevents the system from reaching equilibrium.

## IV. DISCUSSION

The mechanisms of attraction and domain growth that we have proposed are consistent with known experimental trends for DNA condensation. We summarize these trends below.

### A. MINIMUM COUNTERION VALENCE REQUIRED FOR CONDENSATION

DNA generally condenses when trivalent or higher valence cations are added [8]. The charge-fluctuation model predicts that the attraction between rods depends on the cation valence, *Z*, through the variance in the monomer charge, $\Delta \quad Z$, while the repulsion depends on the net charge, *q*, which decreases as *1/Z* . Thus, the net attraction increases rapidly with counterion valence, giving rise to DNA condensation above some threshold $Z_{\min}$. We find that $Z_{\min}$ depends sensitively on the radius of the chains. This is because the repulsion strengthens relative to the attraction as the radius increases, as discussed above. Thus, for larger rod radii, a higher valence is required in order for the attraction to dominate at short distances. As we mentioned earlier, precise microscopic details are important at length scales as short as 6 Å, so we do not expect our results to be quantitatively accurate. However, we do find that the attraction disappears near D=20 Å, characteristic of DNA. Thus, a minimum valence of $Z_{\min}$ =3 is required to condense DNA within our model [48].

### B. EFFECT OF MONOVALENT IONS

When monovalent salt is added at sufficiently high concentrations, it can reverse condensation, leading to isolated chains in solution [11, 51]. This result can also be understood in the context of the charge-fluctuation model. In the model, monovalent salt affects the attraction more strongly than the repulsion because the attraction requires a *correlation* of the charge distributions on different rods. Thus, the first rod must induce a correlated charge distribution on the second rod, and then the two charge distributions must interact. Both steps in this process are screened. The repulsion, on the other hand, merely requires an interaction between the net charges on each rod. As a result, the attraction is screened more heavily than the repulsion. Moreover, at high monovalent salt concentrations, the fraction of condensed monovalent counterions will increase, and the fraction of condensed polyvalent counterions will correspondingly decrease. This will also serve to weaken the attraction relative to the repulsion. At sufficiently high monovalent salt concentrations, the attraction therefore disappears and the bundles fall apart.

One consequence of this argument is that the height of the free energy barrier depicted in Fig. 1 should increase with increasing salt concentration, since the attraction tends to reduce the barrier height. (Above a threshold value, this trend is of course reversed; that is, for sufficiently high salt concentrations, the barrier height eventually decreases with increasing salt concentrations because the attraction is negligible and the repulsion is screened more heavily.) Since the barrier height can increase with salt concentration, this affects the kinetics of domain growth. The barrier height for a rod to join an N-rod bundle is roughly $F_{barrier} \approx \sqrt{N} F_0$, where $F_0$ is the barrier height for two rods to aggregate [47]. The effect of monovalent salt is to increase $F_0$. This means that a smaller aggregation number N is required to achieve a barrier height that is several times the thermal energy kT. As a result, the bundle size should be smaller at higher salt concentrations. This theoretical prediction remains to be tested experimentally.

### C. DEPENDENCE OF BUNDLE THICKNESS ON COUNTERION VALENCE.

The thickness of DNA condensates is found to be roughly independent of *Z* once *Z* exceeds $Z_{\min}$ [8]. In our scenario, the thickness is determined by the height of the barrier depicted in Fig. 1. One would expect the interaction free energy (and therefore the barrier height) to depend sensitively on *Z*. However, we find that the interaction free energy depends only weakly on $Z$ for $Z > 2$ [48]. This may seem counterintuitive at first, but it can be understood simply as follows. Although *condensed* counterions of higher valence induce stronger attractions and weaker repulsions, *free* counterions of higher valence screen electrostatic interactions more effectively. Screening affects the attractions more than the repulsions. Thus, if one is above the threshold concentration of polyvalent counterions required for DNA to condense, the two effects tend to cancel.

## D. DEPENDENCE ON SOLVENT

Experimentally, it is observed that divalent ions can trigger condensation in alcohol-water mixtures. In our model, both $\Delta$ and $q$ also depend on the dielectric constant of the solvent, so that the attraction increases and the repulsion decreases with decreasing $\varepsilon$; this lowers the threshold value of $Z_{min}$. We note, however, that the effect of alcohol could be much more subtle and could depend on microscopic details such as the structure of water near the DNA and counterions that are neglected in our model.

## E. DEPENDENCE ON CHAIN LENGTH

The observed bundle thickness is independent of DNA chain length as long as the length lies in the range between roughly 400 bp (1360 Å) and 50,000 bp [8]. The minimum value of 400 bp could be a buckling length [52] or the length required to make one loop of the toroid [50], and is not explained by our rigid-rod model. However, we do find that the barrier height is independent of chain length as long as the chain length exceeds the persistence length. This is because the appropriate rod length $M$ to choose in order to model the electrostatics of a semiflexible chain is the persistence length, not the chain length.

## F. EFFECT OF BUNDLE SHAPE

The bundle thickness is observed to be roughly the same for both toroidal and rodlike DNA condensates. Several previously suggested mechanisms apply only to toroids [8, 50, 53]. Our mechanism applies to toroidal as well as rodlike bundles. For toroidal bundles, additional chains must overcome the same barrier in order to align with chains already in the torus. Note that our mechanism does not apply to toroids or rods composed of a single chain, because there is no barrier for the unwound part of a chain to join a bundle within our model. However, for chains that are too long (above 50,000 bp), it is difficult to observe bundles because of technical difficulties involved in mixing DNA with polyvalent counterions without breaking the DNA.

## G. TIME DEPENDENCE OF BUNDLE GROWTH KINETICS

Experiments have shown that the initial formation and growth of bundles is quite rapid, occurring over the period of a few minutes. This is followed by very slow growth of bundles over a period of days [51]. These observations are consistent with our picture, where the bundle growth is controlled mainly by diffusion until the barrier height exceeds several kT. Once the barrier height is significant, the growth rate is controlled by the time to cross the barrier, which is long.

### H. OTHER MECHANISMS OF TOROID FORMATION

Recent experiments on the bacteriophage T5 show that arbitrarily large toroidal bundles of condensed DNA can be formed by a special process [54]. In these experiments, the phage DNA is ejected from viruses into liposomes in solution with polyvalent counterions. The amount of DNA ejected into the liposomes can be controlled by increasing the concentration of receptor proteins that enable the viruses to eject their DNA into the liposomes. The kinetics of bundle growth in this experiment are very different from the usual case, because the DNA is released by the virus into the liposome base pair by base pair, in such a way that the orientation of the chain is automatically aligned with those already in the growing toroid. In this way, the system avoids the kinetic barriers discussed in Section III. The fact that the resulting bundles have a much higher cross-sectional diameter lends support to our picture that the bundle size in the usual case is determined by kinetic rather than equilibrium considerations.

## V. SUMMARY

The phenomenon of DNA condensation has spurred physical theorists to address two questions: what is the origin of the polyvalent-counterion-mediated attraction that draws the like-charged chains together into a concentrated bundle, and what sets the characteristic size of the bundle?

The origin of the counterion-mediated attraction between like-charged chains can be understood within the charge fluctuation picture [46]. This picture is conceptually useful because it reconciles the thermal fluctuation approach with the ionic crystal approach. In the thermal fluctuation approach, long-wavelength fluctuations lead to long-ranged attractions that are stronger at high temperatures. In the ionic crystal approach, short-wavelength fluctuations lead to short-ranged attraction that are stronger at lower temperatures. Both mechanisms are captured by the charge-fluctuation approach, as shown for the case of two charged plates [37].

Although most theorists have focused on the *origin* of the counterion-mediated attraction, the main focus of our work has been to explore the *consequences* of the interaction in many-chain systems. The charge-fluctuation approach is particularly well-suited to many-chain systems because it allows an analytical approach [30]. This is particularly important because the counterion-mediated interaction is not pairwise-additive. We have found that, in equilibrium, the system should phase separate on a macroscopic scale into a dilute phase in coexistence with a concentrated phase of parallel chains.

So what sets the characteristic size of a bundle? The charge-fluctuation picture predicts that the system can never reach equilibrium because there is a barrier to bundle growth that increases with the bundle diameter. For a sufficiently large bundle, this barrier is high enough to prevent further growth. These unique domain growth kinetics are a consequence of the unusual angular dependence of the counterion-mediated interaction. Unlike most interactions between anisotropic particles, the counterion-mediated interaction is anisotropic not only in magnitude but also in sign; at short distances, the interaction between two rods is attractive if they are sufficiently parallel, and repulsive otherwise.

Finally, we have shown that the charge-fluctuation approach is consistent with all the trends observed experimentally. Many specific predictions of the model still remain to be tested. We hope that this work will spur further experiments on these fascinating systems.

We thank Robijn Bruinsma and Bill Gelbart for stimulating discussions, and for a careful reading of this manuscript. We gratefully acknowledge the support of the National Science Foundation through Grant No. DMR-9619277.